**Spatial Confinement of the IBEX Ribbon: A Dominant Turbulence Mechanism**

Philip A. Isenberg[1]

[1]Institute for the Study of Earth, Oceans and Space, University of New Hampshire, Durham, NH 03824, USA; phil.isenberg@unh.edu

**Abstract.**   The narrow ribbon of enhanced energetic neutral atom flux observed by the Interstellar Boundary Explorer (*IBEX*) spacecraft has prompted numerous ideas to explain its structure and properties.  One of these ideas is the "neutral solar wind" scenario, which identifies the source particles as pickup protons in the local interstellar medium originating in solar wind charge-exchange interactions.  This scenario has been thought to require unrealistically weak pitch-angle scattering of the pickup protons to explain the narrow structure.  Recently, Schwadron & McComas (2013) suggested that this structure could result from a spatial retention of the pickup protons, rather than from a restricted pitch-angle distribution.  Here, we present a physically motivated, quantitative mechanism to produce such a spatial configuration.  This mechanism is based on the "dominant turbulence" assumption, which can be applied where the production of new pickup protons is slow, and has been used to successfully explain the level of turbulent heating observed in the outer solar wind.  This formalism predicts a pickup isotropization process which adds or subtracts energy from the ambient turbulent fluctuations, depending on the initial pitch angle of the pickup protons.  We show that a simple model of this process can yield a ribbon structure in qualitative agreement with the observations.  The results of this simple model are not yet quantitatively satisfactory, but we suggest several improvements which may reduce the quantitative discrepancy.

## 1. Introduction

Since 2008, the Interstellar Boundary Explorer (IBEX) mission has been collecting data on the properties of the local interstellar medium (ISM) by measuring the fluxes of energetic neutral atoms (ENAs) arriving at Earth orbit.  One of the primary results from this mission is the surprising detection of a ribbon of enhanced ENA flux which circles the heliosphere (McComas et al. 2009; Fuselier et al. 2009; Funsten et al. 2009; Schwadron et al. 2009; McComas et al. 2012).  The flux enhancement over the background is about a factor of 3, confined to a width of $\sim 20°$ and an energy spectrum peaking at $\sim 1 keV$.  Its position in the sky is consistent with the great circle defined by $\hat{\mathbf{r}} \cdot \hat{\mathbf{b}} = 0$, where $\hat{\mathbf{r}}$ is the heliospheric radial direction and $\hat{\mathbf{b}}$ is the direction of the ISM magnetic field according to current best estimates (McComas et al. 2009; Schwadron et al. 2009; Fusilier & Cairns 2013).  The overall structure appears to be steady, but portions of the ribbon exhibit changes in flux over periods of six months or less (McComas et al. 2012), implying that the source of these particles is not too distant.

A number of hypotheses have been advanced to explain the ribbon (see McComas et al. (2009; 2010) or Schwadron et al. (2011) for extensive lists and discussion), but as yet none have convincingly modeled all the observed features.  Perhaps the most popular scenario is the "neutral solar wind" model first proposed by McComas et al. (2009) and further elaborated by Heerikhuisen et al. (2010) and Möbius et al. (2013).  In this picture, the ribbon results from a multi-step process which starts with the well-known pickup of inflowing interstellar hydrogen in the solar wind.  The dominant ionization process in that interaction is charge-exchange with the strongly super-Alfvénic solar wind protons.  This interaction produces a pickup proton tied to the solar wind flow and an energetic hydrogen atom which streams away from the Sun with the velocity of the original solar wind proton.  These neutral solar wind particles flow radially out of the heliosphere and eventually undergo another charge-exchange ionization with the protons in the ISM.  At that point, they have become pickup protons which gyrate about the interstellar magnetic field.

These pickup protons of solar wind origin will behave much like those studied in the solar wind, initially forming a ring-beam distribution given by the energy and pitch angle of the newly ionized particles.  Since these neutrals were streaming radially at the solar wind speed $V_{sw}$, the initial ring-beam has a parallel speed along the interstellar field of $\upsilon_{\|} = V_{sw} \, \hat{\mathbf{r}} \cdot \hat{\mathbf{b}} \equiv V_{sw}$ $\mu_o$, and a gyration speed of $v_\perp = V_{sw}(1 - \mu_o{}^2)^{1/2}$. As in the solar wind, this ring-beam will evolve over time, interacting with ambient and self-generated waves and eventually scattering to a nearly isotropic shell in velocity space. Also over time, these pickup protons are themselves subject to charge-exchange with the hydrogen atoms in the ISM. One of the products of this final charge-exchange will be an energetic neutral atom, again decoupled from the magnetic field, which streams away from the charge-exchange point with the velocity of the former pickup proton. The fraction of these neutrals with instantaneous velocities directed back to the Earth will be detected by the IBEX instruments near the original solar wind energy of ~ 1 keV.

In the models of Heerikhuisen et al. (2010) and Möbius et al. (2013), the narrow image of the ribbon results from assuming that the pickup protons in the ISM do not substantially change their pitch angles between the time of their pickup and the time of their re-neutralization (see also Chalov et al. 2010). In this case, only those protons with $\mu_o \approx 0$ (and the appropriate gyrophase at neutralization) will create neutrals traveling back towards IBEX. Neutrals created from protons with larger $|\mu_o|$ will retain their motion along the magnetic field and proceed in directions transverse to the radial vector, unobserved at Earth.

This last assumption was the principal drawback of this scenario as commonly described. If the pickup protons in the ISM become nearly isotropized before charge-exchanging again, the resulting neutral particles would arrive at IBEX from all directions instead of appearing as a ribbon. The timescale for this final charge-exchange, given the estimated ISM hydrogen density is ~ 2 years (Florinski et al. 2010; McComas et al. 2012). However, these distributions should be unstable (Wu & Davidson 1972; Lee & Ip 1987) and simulations indicate that this narrow energetic ring-beam should undergo significant pitch angle scattering in a much shorter time. Simulations which initially set up the ring-beam with the required density found that it isotropizes on a time of ~ 2 days (Florinski et al. 2010). If the ring-beam is taken to accumulate slowly, as would be the case on a field line approaching the heliosphere from great distance, this time scale may perhaps be extended (Liu et al. 2012), but it is not clear how the required stability could be accomplished in this manner. A study by Gamayunov et al. (2010) has also considered that a combination of external large-scale turbulence and self-generated small-scale turbulence could possibly stabilize the pickup proton distribution without too much spread in pitch angle. However, to the best of our knowledge that preliminary study has not been pursued further.

Recently, Schwadron & McComas (2013, hereinafter SMcC) proposed a different explanation for the ribbon structure within the neutral solar wind scenario. They suggested that the enhanced neutral intensities come from a spatial retention of the ISM pickup protons, rather than due to a pitch angle restriction. They claimed that a pickup proton ring-beam near $\mu_o = 0$ was more unstable to generating waves than otherwise, and that these waves could inhibit the spatial transport of the protons, resulting in a concentration of these particles where $\hat{\mathbf{r}} \cdot \hat{\mathbf{b}} \approx 0$. Although we disagree with some details of the scattering process advanced by SMcC, this paper is important in that it transforms the fundamental question of the neutral solar wind scenario: Rather than looking for a way to strongly inhibit the pitch-angle scattering, perhaps the ribbon could be explained as the result of a localized region of inhibited spatial transport.

Such a spatial confinement requires a physical mechanism which clearly distinguishes conditions leading to strong pitch angle scattering and diffusive particle transport from conditions resulting in weak scattering leaving particles streaming away from the ribbon region. It is this bi-modal character which will determine the resulting spatial structure of the ribbon in these models. The mechanism proposed by SMcC was based on the dispersionless bispherical picture of pickup ion isotropization. According to this picture, pickup protons can only resonate with Alfvén waves propagating in the opposite direction to the proton motion along the magnetic field. Protons with initial streaming speeds less than the Alfvén speed ($|\mu_o| < V_A/V_{sw}$) will scatter to a bispherical shape in velocity space. Such a distribution has a net streaming speed $\approx$ 0, and furthermore is formed through the generation of strong Alfvén waves able to efficiently scatter the particles. The combination of insignificant streaming and strong spatial diffusion in the Alfvén wave field could result in a concentration of pickup protons in this region. In contrast, pickup protons with larger initial streaming speeds in this picture will not be able to scatter through $\mu = 0$, so will continue streaming away from the region of strong diffusion. Following the final charge-exchange neutralization of the pickup protons, a fraction of the ENAs from the diffusive region will be directed toward Earth, but those neutralized outside that region will continue streaming and would not be detected by IBEX.

This is a very appealing scenario, and the SMcC paper presents impressive results comparing their model with observations. However, in our view that model contains a number of questionable details, and should be modified. For instance, the SMcC model invokes strong scattering by unidirectional Alfvén waves propagating toward the center of the retention region, which concentrate the source pickup protons there. At the same time, the pickup protons in the retention region are supposed to be scattered to near isotropy by bi-directional Alfvén waves. Thus, there seems to be an internal inconsistency in this model. Additionally, the bispherical description of pickup ion scattering cited by SMcC was originally formulated for circumstances of rapid ionization rate (Galeev & Sagdeev 1988) and is probably not valid for pickup of the neutral solar wind in the ISM. In a more realistic treatment, the proton scattering by multiple resonant waves would be included. Furthermore, the waves which scatter pickup protons near $\mu$ = 0 are dispersive, and the resulting effects on the scattering process can be important. In this paper, we follow the basic idea suggested by SMcC, but we apply a somewhat different model of the pickup proton scattering process to produce spatial confinement of the ribbon source particles.

The scattering mechanism we propose here is based on the "dominant turbulence" (DT) model of pickup proton isotropization, which has been useful in treating the proton scattering when the ionization of new particles is slow (Isenberg et al. 2003; Isenberg 2005). The application of the DT model to specify the level of turbulent driving in the outer solar wind has been very successful in reproducing the increasing temperature of the core solar wind protons measured at Voyager 2. We point out that a simple extension of this concept can naturally provide a physical mechanism for the bi-modal behavior required by a spatial confinement model. We will show that the incorporation of this mechanism into a rudimentary model of the local ISM can produce a structure appearing as a ribbon of enhanced ENA flux at IBEX. We find that the results of this simple model are qualitatively encouraging, although at this point they do not quantitatively fit the IBEX observations. We will indicate several modifications to be made going forward which may improve these results.

In the next section, we briefly outline the "dominant turbulence" assumption for isotropization of pickup protons and explain how it can lead to a physical criterion for wave generation depending on $\mu_o$. In Section 3, we incorporate the DT mechanism into a simple model of the interaction of the neutral solar wind with the ISM. Section 4 presents our results and discussion. Our conclusions are stated in Section 5.

## 2. Pickup proton isotropization and the "dominant turbulence" assumption

The energetic ring-beam distributions initially formed by newly ionized pickup ions are well-known to be unstable, generating waves which propagate primarily along the magnetic field while the ions scatter to a more isotropic distribution (Wu & Davidson 1972; Lee & Ip 1987). This simultaneous wave generation and pitch-angle scattering is described by the quasilinear resonant cyclotron interaction. This interaction couples ions of parallel speed $v_\parallel$ and gyrofrequency $\Omega = eB/mc$, with particular resonant wave modes of frequency $\omega$ and parallel wavenumber $k$, through the resonance condition

$$\omega(k) - k\,v_\parallel + \Omega = 0. \qquad (1)$$

This resonance condition is illustrated in Figure 1, where we take the ions to be protons and the waves to propagate parallel and anti-parallel to the field according to the cold plasma dispersion relation

$$\omega(k) = \pm k V_A \sqrt{1 + \frac{\omega}{\Omega}}\,. \qquad (2)$$

Here, $V_A$ is the Alfvén speed and both $\omega$ and $k$ may be positive or negative. In this notation, $\omega >$ ($<$) 0 refers to the right-polarized fast mode (left-polarized ion cyclotron mode) waves, and the sign of $\omega/k$ specifies the propagation direction, yielding the four modes $R_\pm$, $L_\pm$ labeled in the Figure. The resonance condition (1) is represented in the Figure by a straight line of slope $v_\parallel$ and $\omega$-intercept at $\omega = -\Omega$. The intersections of such a line with the dispersion curves indicate which waves can resonate with which protons.

We see that protons with small parallel speeds only have one resonance, with either the $L_+$ or $L_-$ wave depending on the sign of $v_\parallel$. Faster protons may have multiple resonances, either with $L_+$ and $R_-$ (for $v_\parallel < 0$) or with $L_-$ and $R_+$ (for $v_\parallel > 0$). From equations (1) and (2), the transition from single to multiple resonances is found at $|v_\parallel| < 3\sqrt{3}\,V_A\,/\,2$.

The form of the fully scattered distribution of pickup protons which results from this resonant interaction depends on the relative intensities of these resonant waves. A commonly assumed form is the "bispherical" distribution (Galeev & Sagdeev 1988; Huddleston et al. 1992; Isenberg & Lee 1996). In the initial scattering of any ring-beam distribution, the resonant mode with one polarization will grow and that of the other polarization will be damped. In the limit that the initial ring-beam of pickup ions is dense enough, one may assume that the damped mode quickly disappears and only the growing mode remains to scatter the ions. When wave dispersion is neglected, the resulting distribution is a shell formed of two spherical sections, each of radius $V_{sw}$, centered on the points $(v_\parallel, v_\perp) = (\pm V_A, 0)$, and joined at the position of the original ring-beam. In rapidly ionizing environments such as found upstream of comets, the bispherical assumption is justified and useful.

However, in circumstances where the ionization and pickup of new ions is slow due to reduced ionization rates or low neutral density, the "damped" modes do not disappear and one cannot neglect the effects of scattering by these waves. An important example of this slower process is found for the case of turbulent heating of the core protons in the distant solar wind. The temperature of the core protons measured by Voyager 2 does not continually decrease with heliocentric radial position, but rather levels off and starts increasing beyond ~ 40 AU (Richardson et al. 1995; Richardson & Smith 2003). Williams et al. (1995) first suggested that the waves generated by pickup and isotropization of inflowing interstellar hydrogen could provide an additional energy source to the solar wind turbulence, which could then lead to these increasing temperatures from the additional turbulent dissipation. The turbulent driving from pickup isotropization is obtained by taking the fully scattered shape of the pickup distribution and comparing the energy of that shell of protons to that of the original ring-beam. The energy difference in the particles goes into the waves and this wave energy is added to the turbulent intensity. In the early studies of this heating model, the fully scattered proton distribution was taken to be the bispherical result. The initial ring-beam of pickup protons in the azimuthal magnetic field of the distant equatorial solar wind will have $\mu_o = 0$, so the appropriate bispherical distribution is a symmetric football-shaped shell produced by interaction with the $L$-mode waves only. This fully scattered bisphere has a final energy less than the initial ring-beam, and channels a fraction on the order of $V_A/V_{sw}$ of the initial energy into the turbulence. It was found, however, that this assumption yielded a turbulent driving that was far too strong, giving much higher temperatures than those measured at Voyager (Smith et al. 2001).

In the outer solar wind, though, the time scale for ionization of new pickup protons is several months to one year. With such a slow input, there is ample time for the turbulent redistribution of the spectral power in resonantly growing or decaying waves, invalidating the underlying justification for the bispherical assumption. That said, it is not immediately clear how to choose a better description for the pickup proton isotropization. Turbulent cascades in collisionless magnetized plasmas primarily transport fluctuation energy in the direction perpendicular to the large-scale magnetic field (Shebalin et al. 1983; Montgomery & Turner 1981; Goldreich & Sridhar 1995), so research in this area has concentrated on the nonlinear properties of these modes. However, highly perpendicular fluctuations do not effectively scatter energetic ions toward isotropy (Chandran 2000; Shalchi & Schlickeiser 2004) and observations in the solar wind have consistently found evidence for a component of the turbulent power in quasi-parallel fluctuations (Matthaeus & Goldstein 1982; Bieber et al. 1996; Dasso et al. 2005). Within the context of the quasilinear resonant cyclotron interaction, the failure of the bispherical approximation in the outer solar wind strongly implies that a substantial fraction of the unstable *L*-mode wave power must be transferred into *R*-mode fluctuations. A simultaneous interaction with both *L*-mode and *R*-mode waves would scatter the pickup protons into a more energetic, and more isotropic, shell.

To investigate this possibility, the "dominant turbulence" (DT) assumption was introduced, taking the spectral redistribution of quasi-parallel waves to be rapid compared to the time scale for pickup of new particles (Isenberg et al. 2003; Isenberg 2005). Under this assumption, we take the resonant wave spectrum to be maintained as a power law in *k* with equal intensities in the four resonant modes:

$$I_{\pm}(k) \sim |k|^{-5/3}, \qquad (3)$$

where the ± refers to the sign of $\omega$, and *k* may be positive or negative. This assumption leads to a different form for the fully scattered distribution, one which is more isotropic and provides less energy to the fluctuations than the bispherical shape. In a number of papers, with increasing levels of detail, this assumption has provided an energy input to the turbulence that leads to good agreement with the temperatures measured at Voyager 2 (Smith et al. 2006; Isenberg et al. 2010a; Isenberg et al. 2010b; Oughton et al. 2011).

The quasilinear analysis behind this fully scattered distribution is derived in Isenberg (2005), so we simply list the relevant equations here. (They are equations (12), (13), and (17) of that paper.) At any position along the fully scattered shell $v(\mu)$, the local shape is given by the relative intensities of the waves resonant with these particles, as

$$\frac{dv}{d\mu} = \left[ \sum_j V_j \frac{I_j(k_r)}{|\mu v - W_j|} \left( 1 - \frac{\mu V_j}{v} \right) \right] \left[ \sum_j \frac{I_j(k_r)}{|\mu v - W_j|} \left( 1 - \frac{\mu V_j}{v} \right)^2 \right]^{-1}, \qquad (4)$$

where the sums are taken over all the cyclotron resonances at that value of $v_\parallel = \mu v$. The functions $V_j$ and $W_j$ are the resonant wave phase and group speeds, respectively, obtained from (2), and the individual resonant wavenumbers are given by (1) as

$$k_r = \frac{\Omega}{\mu v - V_j}. \qquad (5)$$

With the relative intensities given by (3), equation (4) can be integrated to give the fully scattered distribution, which only depends on the value of $V_A/V_{sw}$ and the initial pitch angle of the ring-beam. (The absolute wave intensity disappears from the ratio (4). Physically, it sets the time scale for the isotropization, which we take to be small under the DT assumption.)

Some examples of fully scattered shells are shown in Figure 2 for several values of $\mu_o$ and for $V_A/V_{sw} = 1/18$ (appropriate to pickup of neutral solar wind moving at 450 km/s in an ISM with Alfvén speed of 25 km/s). The two-part structure of each curve is due to the change in the resonance condition at $\mu_c = 1.5\sqrt{3}\, V_A / v$. For $|\mu| < \mu_c$ the protons can only resonate with the $L$ mode, as in the bispherical case when $\mu_o = 0$. The equivalent bispherical shape would be the continuation of the small-$\mu$ curve for all $\mu$. However, for $|\mu| > \mu_c$ the protons are also scattered by the $R$ modes, which we take to be maintained by the turbulent interactions. The combined scattering by equal intensity $L$ and $R$ modes results in a much more isotropic shape and much less energy lost by the protons. For this small value of $V_A/V_{sw}$, the anisotropies of the fully scattered distributions are on the order of 2%, almost three times smaller than the equivalent bispherical result. We also see that the shapes of the various curves in Figure 2 are quite similar. The major difference in these curves is the normalization, defined by the condition $v(\mu_o) = V_{sw}$.

The net fluctuation energy added to the existing turbulence by this isotropization is equal to the energy lost by the particles in scattering from the ring-beam at $E_o = 0.5\, m V_{sw}^2$ to the fully scattered distribution. As detailed in Isenberg (2005), we numerically obtain the energy of this final distribution by calculating velocity moments of the constant-density shells. The shells defined by (4) do not have constant thickness as functions of $\mu$, so a scale factor, $S(\mu)$, must be included in the moment integrals. Thus, the fraction of the initial energy which is available for turbulent driving is

$$\zeta\left(\mu_o\right) \;=\; 1 - \frac{\int_0^1 v^4\left(\mu\right)S(\mu)d\mu}{V_{sw}^2\int_0^1 v^2\left(\mu\right)S(\mu)d\mu} \;. \tag{6}$$

We obtain the scale factor $S(\mu)$ by integrating (4) with the condition $v\left(\mu_o\right) = V_{sw}$, and then again with $v\left(\mu_o\right) = 1.001\,V_{sw}$, taking the difference at each $\mu$ normalized to unity at $\mu = \mu_o$. The fractional energy loss from (6) for $V_A/V_{sw} = 1/18$ is shown by the black curve in Figure 3.

We see from Figure 3 that $\zeta$ changes sign for intermediate values of $|\mu_o|$ (specifically at $|\mu_o| = 0.137, 0.156$, and $0.406$). A ring-beam at small $\mu_o$ will be scattered by the turbulent spectra (3) to a distribution of slightly lower energy, and the scattering process will add this lost energy to the fluctuations. However, if the ring-beam starts at $|\mu_o| > 0.406$, the scattering process will require additional energy to complete, and it will extract that energy from the ambient fluctuations, to the extent that sufficient wave intensity is present. Conversely, if the fluctuation level is too small, the waves will not be able to scatter the new protons over the energy maximum at $\mu = 0$, so the protons picked up at these larger values of $|\mu_o|$ will not be able to isotropize.

As noted by SMcC, quasilinear theory relates the level of resonant fluctuations to the spatial diffusion rate of the particles. A spatially limited region of high resonant intensities can confine a nearly isotropic distribution of pickup protons to that region. We have now seen how the DT scattering of newly picked up protons can enhance the turbulent intensity when the pitch angle cosine of the initial ring-beam is small. The initial pitch angle cosine $\mu_o = \hat{\mathbf{r}} \cdot \hat{\mathbf{b}}$, so the region of enhanced turbulence will be generated where $\hat{\mathbf{r}} \cdot \hat{\mathbf{b}} \approx 0$, exactly as required for a ribbon model. Thus, the DT assumption provides a physically motivated mechanism for concentrating the pickup proton density in a region consistent with the source of ribbon ENAs.

In the next section, we will incorporate a form of this DT isotropization into a model of the local ISM to address the feasibility of spatially confining pickup protons into a ribbon source.

### 3. Spatial Confinement Model

In this section, we construct a rudimentary model of the local ISM, closely following the valuable work of Möbius et al. (2013). We assume that the background ISM parameters - the plasma density $n_p$, the hydrogen density $N_H$, the flow velocity $\mathbf{V_{ISM}}$, and the magnetic field $\mathbf{B_{ISM}}$ - are all constant and are unaffected by flow deflection or field-line draping around the heliosphere. We label the angle between $\mathbf{V_{ISM}}$ and $\mathbf{B_{ISM}}$ as $\theta_{BV}$ = constant. For this paper, we also limit our calculations to a single plane, shown in Figure 4. This plane is defined by the ISM flow velocity, the ISM field direction, and the position of IBEX, taken to be indistinguishable on these scales from the position of the Sun.

In this plane, we seek a steady-state solution for the coupled system of nearly isotropic pickup protons of solar origin and the spectrum of turbulent fluctuations which interacts with them. We define $n(x, y)$ as the density of those pickup protons which originate as neutral solar wind and which scatter to near-isotropy in the ISM. These protons have been ionized in regions of space where the self-generated or pre-existing turbulence is strong enough to rapidly isotropize the initial ring-beam. This fully scattered proton population will accumulate in the ISM plasma and diffuse along the magnetic field as the plasma flows toward the heliosphere. It is only these protons which have a chance of heading back to the Earth as ~ 1 keV ribbon ENAs when they become neutralized through charge-exchange with interstellar hydrogen.

The fluctuation intensity is defined in the Elsässer form by $Z^2 = <\delta v^2> + <\delta b^2/4\pi\rho>$, where $\delta v$ and $\delta b$ are the velocity and magnetic field fluctuation amplitudes, respectively, $\rho$ is the plasma density, and the angular brackets denote an ensemble average. The fluctuations will be driven by the isotropization of newly ionized protons and will mediate their spatial diffusion along the magnetic field. The fluctuation intensity will also be advected by the ISM motion toward the heliosphere, while continuing to be amplified or absorbed by the subsequent pickup proton scattering.

In other regions of space, the previously generated turbulent intensity will be too small to effectively isotropize the newly ionized protons. In this case, the locally ionized protons will continue streaming along the magnetic field at some fraction of $V_{sw}$ in directions away from the $\mu_o \approx 0$ region. They will not be confined by diffusive scattering and will not be considered further in this paper.

The diffusive aspect of this model leads to a different form of particle transport than is found in earlier neutral solar wind models of the ribbon. Models which take the pickup protons to be essentially scatter-free (e.g. Chalov et al. 2010; Heerikhuisen et al. 2010; Möbius et al. 2013) need only couple these particles to the component of the ISM motion perpendicular to the magnetic field. In this case, the pickup population is advected toward the Sun at a speed $u_o = V_{ISM} \sin(\theta_{BV})$, and the protons picked up at $\mu_o = 0$ will remain on the line labeled "**y**" shown in Figure 4. In contrast, we want to treat the diffusive behavior of a nearly isotropic proton distribution, which therefore will be strongly coupled to the local ISM plasma, and will move at $\mathbf{V}_{ISM}$ (horizontally in Figure 4) independent of the angle of **B**. At the same time, the pitch-angle cosine of the initial ring-beam of pickup protons still depends on the field angle. This means that the pickup protons initially created at $\mu_o = 0$ will start on the line **y** but will be displaced to the left in the plane of Figure 4 as they are advected closer to the Sun. Thus, we expect the region of enhanced turbulence and source of ribbon ENAs to be slightly skewed, as a function of distance, in the direction away from the nose of the heliosphere.

Since we need to track the pickup proton density as it follows a plasma parcel in the ISM and diffuses along the magnetic field, we use a coordinate system appropriate to that motion. We take a Cartesian system with the $x$-axis parallel to $\hat{\mathbf{b}}$ and the $y$-axis pointing away from the Sun along **y**, as shown in Figure 4. The $x = 0$ point follows $\mathbf{V}_{ISM}$ and intersects the $y$-axis at the tangent point of the magnetic field with the heliopause, as shown by the dashed line in the Figure. The coordinates of a plasma parcel advected toward the heliosphere will have $x =$ constant, while $y$ will decrease as $y =$ constant $- u_o t$. In this system, the value of $x$ on the $y$-axis is a function of $y$, given by $x_r(y) = (y - r_{HP}) \cot \theta_{BV}$ where $r_{HP}$ is the heliocentric position of the heliopause. At any point, the radial distance from the Sun is then $r(x, y) = [y^2 + (x - x_r)^2]^{1/2}$.

To model the macrophysics of the ribbon in a simple, but still meaningful way, we make several further approximations in the microphysics of the DT interaction as applied to the ISM. For the remainder of this paper, we will treat the behavior of the nearly-isotropic fully scattered distribution as though it is truly isotropic, in a shell at $\upsilon = V_{sw}$. In this sense, we also neglect the small contribution of $V_{ISM} << V_{sw}$ to the products of the charge-exchange interactions. We further simplify the detailed energy partition function, $\zeta(\mu_o)$ in equation (6) so that it falls monotonically with increasing $\mu_o$. In clearly defining a single range of $\mu_o$ where pickup isotropization causes wave growth rather than wave damping, we avoid some messy behavior of this simple model near the spatial boundaries of the isotropized particles. This modification is accomplished by replacing the intermediate portion of $\zeta$ by a linear segment, shown by the red line in Figure 3. This composite function will be used in the remainder of this paper. Finally, from this point on we will use the simplified expressions for the quasilinear proton diffusion coefficient that result when $V_{sw} >> V_A$ and dispersion of the resonant waves is ignored.

Proceeding under these assumptions, we start with the neutral solar wind which streams radially away from the Sun at speed $V_{sw}$. The charge-exchange rate is given by the cross-section for this process, $\sigma$, times the local densities of the neutral and ionized hydrogen multiplied by the relative speed between them. Thus, the neutral solar wind particles will be ionized and picked up at a production rate $P = \sigma N_{sw} n_p V_{sw}$, where $N_{sw}$ is the local density of the neutral solar wind. Still following Möbius et al. (2013), we assume that the neutral solar wind density expands spherically from a normalized value, $N_{sw}^o$, at a spherical termination shock set at $r_o = 100$ AU, and is not affected by the plasma in the inner heliosheath. Beyond the heliopause, however, the neutral solar wind density is attenuated by charge-exchange on a length scale equal to $(\sigma n_p)^{-1}$, so the pickup proton production rate is

$$P(x,y) = \sigma n_p V_{sw} N_{sw}^o \frac{r_o^2}{r^2} \exp\left[-\sigma n_p\left(r - r_{HP}\right)\right]$$

$$= \sigma n_p V_{sw} N_{sw}^o \frac{r_o^2}{y^2 + (x - x_r)^2} \exp\left[-\sigma n_p\left(\sqrt{y^2 + (x - x_r)^2} - r_{HP}\right)\right]. \qquad (7)$$

At the same time, those particles which have already been picked up in the ISM are subject to loss by charge-exchange at a rate

$$L = \sigma N_H V_{sw} n(x,y). \qquad (8)$$

The pitch angle of a newly ionized neutral solar wind particle in our coordinate system is given by

$$\alpha = \tan^{-1}\left(\frac{y}{x - x_r}\right), \qquad (9)$$

so the pitch angle cosine of the initial ring-beam in this plane is

$$\mu_o = \frac{x - x_r}{\sqrt{y^2 + (x - x_r)^2}} \ . \tag{10}$$

In those regions where the turbulent fluctuations are large enough to isotropize the distribution, the pickup protons will diffuse along the magnetic field with a diffusion coefficient $D(x, y)$. Thus, in steady-state, the ISM density of isotropic pickup protons originating in the neutral solar wind is given by the solution of

$$-u_o \frac{\partial n}{\partial y} = \frac{\partial}{\partial x} \left( D \frac{\partial n}{\partial x} \right) + P - L \ . \tag{11}$$

In this equation, the left-hand side is due to the advection of the protons on a magnetic field line with the ISM plasma speed. The terms on the right-hand side denote the spatial diffusion along the magnetic field, the production of new protons (7), and the loss of protons to charge-exchange (8), respectively. The quasilinear spatial diffusion coefficient is inversely proportional to the pitch-angle scattering rate, $D_{\mu\mu}$, as (Hasselmann & Wibberenz 1970)

$$D = \frac{V_{sw}^2}{4} \int_{-1}^{1} \left[ \int_0^{\mu'} \frac{1 - \mu^2}{D_{\mu\mu}} \, d\mu \right] \mu' d\mu' \ . \tag{12}$$

Under the DT assumption, the resonant fluctuations are treated as components of a turbulent spectrum, which efficiently re-distributes their intensities. Here, we extend the assumption of (3) to include a flattening of the power spectrum below a correlation wavenumber $k_o$ in order to obtain an absolute intensity in the model:

$$I_{\pm}(k) = \begin{cases} A \left( \dfrac{|k|}{k_o} \right)^{-5/3} & |k| > k_o \\ A & |k| < k_o \end{cases} \ . \tag{13}$$

With this spectrum and the limit that the particle speed $V_{sw} >> V_A$, the quasilinear pitch-angle diffusion coefficient is (Lee 1971; Schlickeiser 1989)

$$D_{\mu\mu} = \frac{\pi}{2} \left( \frac{e}{mc} \right)^2 \left( 1 - \mu^2 \right) \sum_j \frac{I(k_j)}{|\mu V_{sw}|} \tag{14}$$

where the sum is over the resonances $k_j = \pm \Omega/|\mu V_{sw}|$ and we assume $|k_j| > k_o$. (Recall that in this section, we ignore the effects of wave dispersion.) A given pickup proton will resonate with two of the equal-intensity modes of (13), so

$$D_{\mu\mu} = \pi \frac{A}{B^2} \left(1 - \mu^2\right) \frac{\Omega^2}{|\mu| V_{sw}} \left(\frac{\Omega}{|\mu| k_o V_{sw}}\right)^{-5/3} \tag{15}$$

and the spatial diffusion coefficient (12) is then

$$D = \frac{9 V_{sw}^3}{14\pi} \frac{B^2}{A\Omega^2} \left(\frac{\Omega}{k_o V_{sw}}\right)^{5/3} . \tag{16}$$

In this model, we also need to treat the turbulent evolution, which we represent in the simplest possible manner. We do not try to model the nonlinear cascade physics here. Instead, we base our treatment on the phenomenological formalism of von Kármán & Howarth (von Kármán & Howarth 1938; Matthaeus et al. 1996; Zhou & Matthaeus 1990; Zank et al. 1996; Oughton et al. 2006; Breech et al. 2008; Oughton et al. 2011), which considers the evolution of the fluctuation intensity at the energy-containing scales. For the purposes of this model, we track only the quasi-parallel component of the turbulence, since these modes are directly resonant with the protons (Chandran 2000; Shalchi & Schlickeiser 2004), and so control both the pitch-angle scattering and the spatial diffusion. This treatment may be thought of as a reduction of the recent two-component model of Oughton et al. (2011) in the limit that their coupling between the quasi-parallel and quasi-perpendicular fluctuations goes to zero and all correlation lengths are taken equal. In this picture, we describe the relevant turbulence by the ensemble-averaged intensity in Elsässer units, $Z^2$ (defined above) and a correlation length, $\lambda_\parallel$.

Within our model for pickup of neutral solar wind in the ISM, the turbulent intensity will increase or decrease due to the input from the pickup proton isotropization and will also decay at an Iroshnikov-Kraichnan dissipation rate $\approx Z^4/(\lambda_\parallel V_A)$. In a steady state system with a homogeneous background, the turbulent intensity equation following a fluid parcel in the ISM is then (see e.g. Isenberg 2005, equations (1) and (4); Oughton et al. 2011, equation (8))

$$-u_o \frac{\partial Z^2}{\partial y} = -\frac{2Z^4}{\lambda_\parallel V_A} + Q , \tag{17}$$

where $Q$ is the turbulent generation rate from the pickup protons under the DT approximation

$$Q = \zeta(\mu_o) \frac{V_{sw}^2}{n_p} P(x,y) . \tag{18}$$

Here, $Q$ is the product of the production rate of new protons (7) and the fractional energy input function shown in Figure 3, written in Elsässer units.

The correlation length evolves as well, tending to increase as the turbulence cascades toward dissipation and also tending to decrease due to the pickup proton driving at their resonant scale $\lambda_{res} = 2\pi V_{sw}/\Omega$. This equation can be written (e.g. Oughton et al. 2011, equation (13))

$$-u_o \frac{\partial \lambda_\parallel}{\partial y} = \frac{2Z^2}{V_A} - (\lambda_\parallel - \lambda_{res})\frac{Q}{Z^2} S(Q).$$ (19)

In comparison to previous applications of this formalism, we have added a step function of $Q$, denoted by $S(Q)$, to the last term in (19). Without this modification, circumstances of $Q < 0$ and small $Z^2$ lead to an unrealistic dominance by this term and a strongly increasing $\lambda_\parallel$ at the boundaries of the diffusive region. Here, we suggest that the additional damping of the intensity represented by $Q < 0$ in (17) appears to the turbulent spectrum as simply additional high-$k$ dissipation with negligible further effect on $\lambda_\parallel$.

Finally, we relate the turbulent intensity to the resonant spectrum (13) through the definition of the total magnetic fluctuation energy,

$$<\delta b^2> \equiv \sum_\pm \int_{-\infty}^{\infty} I_\pm(k)dk = 10Ak_o,$$ (20)

so

$$\frac{B^2}{Ak_o} = 10(1 + r_A)\frac{V_A^2}{Z^2}$$ (21)

where $r_A$ is the Alfvén ratio, $4\pi\rho<\delta\upsilon^2>/<\delta b^2>$, taken to be constant, $r_A = 1/2$ (Roberts et al. 1987). Identifying the correlation length $\lambda_\parallel = 2\pi/k_o$, the spatial diffusion coefficient in (16) becomes

$$D = \frac{45}{7\pi(2\pi)^{2/3}}(1 + r_A)V_{sw}^{4/3}V_A^2\Omega^{-1/3}\frac{\lambda_\parallel^{2/3}}{Z^2}.$$ (22)

With these expressions, we see that the diffusion coefficient can be very large if $Z^2$ is small. However, there must be a value of $Z^2$ which is too small to isotropize the initial ring-beam of pickup protons rapidly enough to be included in the model ribbon. We estimate this value by setting the characteristic pitch-angle scattering rate equal to the production rate for new pickup protons from the neutral solar wind. To a reasonable approximation, this equality can be written

$$\frac{<\delta b^2>}{B^2}\Omega \approx \sigma n_p V_{sw}.$$ (23)

Thus, for $Z^2$ to provide enough scattering to isotropize the pickup distribution in the face of a continual source of new particles at $\mu_o$, it must satisfy

$$Z^2 > (1 + r_A)\frac{\sigma n_p}{\Omega}V_A^2 V_{sw} \equiv g_{min}.$$ (24)

In regions of the model where $Z^2 < g_{min}$, we take the newly picked up protons to remain close enough to their initial ring-beam that they stream away from the DT region, and we set $n = 0$ there. We will see that the model results do not depend sensitively on the value of $g_{min}$.

The condition (24) is a considerable idealization, though in keeping with the simplified nature of this model. Of course, the transition from isotropic, diffusively confined pickup protons to weakly scattered streaming protons will not be abrupt in reality. A more detailed and rigorous treatment of this transition layer could perhaps lead to useful improvements of this model.

The solution of the simultaneous equations (11), (17) and (19) as a field line at distant $y$ moves closer to the heliosphere will provide the spatial distribution of nearly isotropic pickup protons due to the neutral solar wind in the model plane. For the results presented in the next section, we take the physical quantities in the ISM to be the constant values shown in Table 1, where the values below the double line are derived from those above. We further assume that the ribbon is a local, self-generated phenomenon, so we take the conditions at the outer boundary, $y = y_o = 2100$ AU, to be $n(y_o) = Z^2(y_o) = 0$, and $\lambda_{\parallel}(y_o) = \lambda_{res} = 6.5 \times 10^{-4}$ AU. We set the side boundaries of our system at $x = -500$ AU and $x = 2500$ AU, and also require $n = 0$ there. The equations are first-order in $y$ and we integrate them inward, toward the Sun. At each numerical step in $y$ we first solve for the advanced turbulent quantities, $Z^2$ and $\lambda_{\parallel}$, with an explicit predictor-corrector scheme. If the numerical solution for the new $Z^2$ yields $Z^2 < 0$, it is reset to $Z^2 = 0$. Using the previous and advanced values of $Z^2$ and $\lambda_{\parallel}$, the diffusive pickup density, $n$, is then calculated along the new field line with an implicit Crank-Nicholson scheme. The equations are solved using $\Delta x = \Delta y = 0.5$ AU. We set the heliopause at $r_{HP} = 150$ AU, but continue the numerical integration inward to $y = 100$ AU in order to obtain solution values on the flanks of the heliopause.

## 4. Results and Discussion

In Figures 5 - 7, we show the results of this calculation in the model plane, transformed into heliocentric $r$, $\theta$ coordinates, where $\theta = 0$ is taken in the $\hat{\mathbf{r}} \cdot \hat{\mathbf{b}} = 0$ direction. The figures display contours of $\log(n)$, $\log(Z^2)$ and $\log(\lambda_{\parallel})$, respectively, for $r \geq 150$ AU, where $n$ is in cm$^{-3}$, $Z^2$ is in (km/s)$^2$, and $\lambda_{\parallel}$ is in AU. The correlation length results are only plotted for positions where $Z^2 > 0$. We see that the self-generated turbulence is limited to a region around the $\theta = 0$ axis, and that the fluctuations diffusively confine the nearly isotropic pickup protons to an equivalent region. As expected, the pickup proton density is slightly skewed to positive $\theta$ (in the direction of the ISM flow). The pickup density is also concentrated near the heliopause, giving a natural explanation for the time variations observed in the ribbon.

As stated above, this model sets the density of isotropic pickup protons equal to zero where the turbulent intensity falls below a cutoff value of $g_{min}$. The results in Figure 6 show that $Z^2$ falls off very steeply on the flanks of the model ribbon, so we conclude that the boundaries of the $n > 0$ region will not move very much if the value taken for $g_{min}$ is changed. Thus, the shape of the model ribbon is not sensitive to the details of the transition from isotropic, diffusive behavior to weakly scattered, streaming behavior.

The final step in a ribbon model is to calculate the ENA intensity which would be measured at IBEX. The nearly isotropic pickup protons in this model solution charge-exchange again with the neutral hydrogen in the ISM, radiating ~1 keV ENAs in all directions. Those ENAs directed toward Earth will produce a differential flux at the heliopause from the $\theta$-direction of

$$J(r_{HP}, \theta) = \frac{\sigma N_H V_{sw}}{4\pi \Delta E} \int_{r_{HP}}^{\infty} n(r, \theta) \exp\left[-\sigma n_p (r - r_{HP})\right] dr \,, \tag{25}$$

where we also take into account the loss of these ENAs to further charge-exchange along their path. This ENA flux is plotted in Figure 8 for an IBEX energy resolution of $\Delta E = 0.7$ keV.

As a representation of a cut through the IBEX ribbon, this result is qualitatively encouraging, though the quantitative comparison with the observed IBEX fluxes is not yet satisfactory. The enhanced flux in Figure 8 is broader than observed, and the maximum of $J =$ 22.4 (cm$^2$ sr s keV)$^{-1}$ is a factor of 8 - 10 too low (Schwadron et al. 2011; McComas et al. 2012).

It is interesting to compare our results to those of Möbius et al. (2013), who were able to reproduce the observed ribbon flux at the heliopause by assuming no pitch-angle scattering of the ISM pickup protons. They emphasized that, if the pickup protons in their calculation were, instead, taken to be isotropic, the ENA flux from their model would fall to an insufficient value (along with negating the explanation for the localized ribbon structure of course). In fact, our peak ENA flux is comparable to that obtained by Möbius et al. in their isotropic case. This indicates that our mechanism, while creating a localized ribbon in the presence of the expected pitch-angle scattering, has yet to produce the required concentration of pickup protons in this region.

In contrast, the results obtained by SMcC showed fluxes much more in agreement with the IBEX observations of the ribbon. In their model, the strong concentration of pickup protons within the ribbon structure is accomplished by, in effect, an additional scattering by unidirectional Alfvén waves. These waves are taken to propagate toward the center of the ribbon from both sides and thus provide a pile-up of the source pickup protons. We doubt that such an imbalanced wave intensity exists in the ISM, but some form of active concentration must be taking place if the spatial retention scenario is to provide a viable explanation of the ribbon.

We are currently exploring how this concentration might be accomplished within our dominant turbulence framework. The simple model presented here is missing a number of effects which could be important in this regard. For instance, we have not included the deflection of the ISM flow or the draping of the magnetic field as the plasma approaches the heliosphere (Chalov et al. 2010). There could also be significant modifications of the quasi-parallel resonant wave spectra by the effect of the turbulent transport of wave power to smaller perpendicular scales (Oughton et al. 2006; Oughton et al. 2011). A more realistic treatment of the transition at the edges of the diffusive region from isotropic to streaming pickup protons could also perhaps lead to an increased proton density in the ribbon. We further suspect that including the contribution of additional charge-exchanges from the ENAs emitted by the confined pickup protons could lead to a further concentration of the pickup density in the diffusive region.

## 5. Summary and Conclusions

We have presented a model for the IBEX ribbon of enhanced ENA flux within the context of the neutral solar wind scenario (McComas et al. 2009; Heerikhuisen et al. 2010; Möbius et al. 2013). This model does not assume unrealistically weak pitch-angle scattering of the ISM pickup protons, but follows the suggestion of Schwadron & McComas (2013) who proposed a spatial confinement explanation for the ribbon. Specifically, we have developed a new, physically motivated mechanism which localizes the ribbon source particles due to a bi-modal behavior of the pickup proton isotropization process consistent with quasilinear theory. This "dominant turbulence" mechanism invokes efficient turbulent redistribution of wave power when production of new pickup protons is slow. This mechanism has been successfully applied to explain the level of core proton heating observed in the outer solar wind by Voyager 2. We evaluated the predictions of this ribbon model for a simple ISM system in the single plane containing the ISM flow, ISM magnetic field and the Sun. We found that this model can produce a localized ribbon structure qualitatively similar to that observed by IBEX. However, the quantitative comparison with observations is not yet satisfactory, yielding an enhanced flux of ENAs which is smaller and broader than that seen by IBEX. In future work, we will explore improvements to this model, which may reduce this discrepancy.

**Acknowledgements.** The author is grateful for valuable conversations and advice from M. A. Lee, D. J. McComas, E. Möbius, N. A. Schwadron, C. W. Smith, B. J. Vasquez, and D. K. Isenberg. This work was supported in part by NASA grants NNX13AF97G and NNX11AJ37G, and NSF grant AGS0962506.

**Table 1.** Model ISM Parameter Values

| | |
|---|---|
| $n_p$ (cm$^{-3}$) ................................. | 0.07 |
| $N_H$ (cm$^{-3}$) ................................. | 0.16 |
| $V_{ISM}$ (km s$^{-1}$) ........................ | 23.2 |
| $V_{sw}$ (km s$^{-1}$) ......................... | 450. |
| $V_A$ (km s$^{-1}$) ............................. | 25. |
| $\theta_{BV}$ (deg) ................................ | 45 |
| $r_{HP}$ (AU) .................................... | 150. |
| $\sigma$ (cm$^2$) ................................... | $2. \times 10^{-15}$ |
| $r_A$ ............................................. | 0.5 |
| $N_{sw}{}^o$ (cm$^{-3}$) ............................ | $8.2 \times 10^{-5}$ |
| $u_o$ (km s$^{-1}$) ............................. | 16.4 |
| $B_{ISM}$ ($\mu$G) ......................... | 3.03 |
| $\Omega$ (rad s$^{-1}$) ............................ | $2.9 \times 10^{-2}$ |
| $\lambda_{res}$ (AU) ............................... | $6.5 \times 10^{-4}$ |
| $g_{min}$ (km$^2$ s$^{-2}$) ......................... | $2.04 \times 10^{-4}$ |

**Figure Captions**

**Figure 1.**  Representation of the cyclotron resonance condition between protons and dispersive parallel-propagating MHD waves.  The solid curves show the dispersion relation in the $\omega$-$k$ plane for the four possible modes: fast ($R$) and ion-cyclotron ($L$) each propagating in the positive or negative direction along the magnetic field.  A proton will be resonant with one of these modes if a straight line passing through $(\omega, k) = (-\Omega, 0)$ with slope equal to the proton parallel speed $\mu v$ intersects the dispersion curve of that mode.  The dashed line illustrates the small parallel speed case, $\mu < \mu_c$, which can only resonate with one mode.  The solid straight line illustrates the large parallel speed case, $\mu > \mu_c$, which allows three resonances (the high frequency $R$ resonance is not shown).

**Figure 2.**  Shapes of fully scattered pickup proton shells for $V_A/V_{SW} = 1/18$, under the DT assumption for four values of the initial pitch angle $\mu_o$.  The shapes are symmetric about $\mu = 0$.

**Figure 3.**  Fraction of the initial pickup proton energy which is transferred into wave energy as a result of pitch-angle scattering under the DT assumption.  This fraction maximizes for initial pitch angle $\mu_o = 0$.  Negative values represent conditions which take energy from the ambient waves to complete the scattering.  The black curve is the detailed solution of equation (6), and the straight red segment shows the simplification used in this paper.

**Figure 4.**  Spatial configuration of our ISM model in the plane containing the ISM flow vector, $\mathbf{V_{ISM}}$, the ISM magnetic field vector, $\mathbf{B_{ISM}}$, and the Sun.  Other quantities are explained in the text.

**Figure 5.**  Logarithmic contours of pickup proton density, $n$, in the model plane using heliocentric polar coordinates.  The density has units of $cm^{-3}$.

**Figure 6.**  Logarithmic contours of turbulent intensity, $Z^2$, as in Fig. 5.  The intensity has units of $(km/s)^2$.

**Figure 7.** Logarithmic contours of the turbulent correlation length, $\lambda_{\parallel}$, as in Fig. 5. The length has units of AU, and values are only plotted where $Z^2 > 0$.

**Figure 8.** Predicted flux of ribbon ENAs in the model plane at the heliopause, $r = 150$ AU.

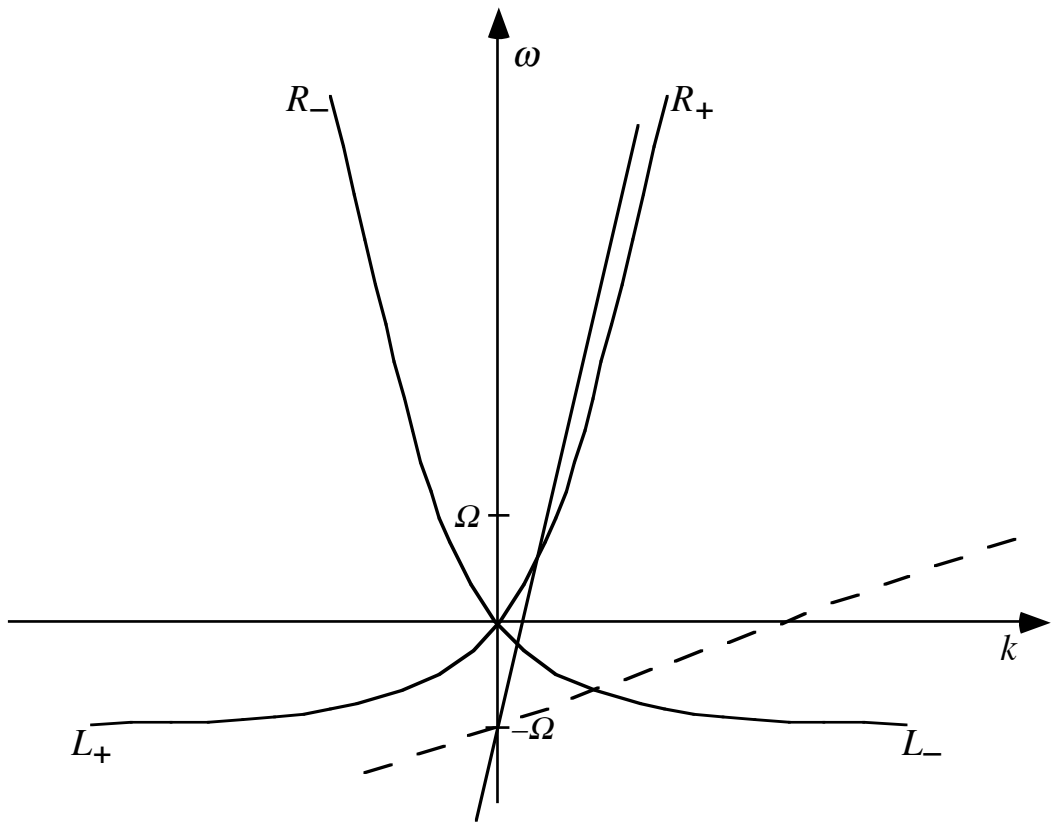

Figure 1

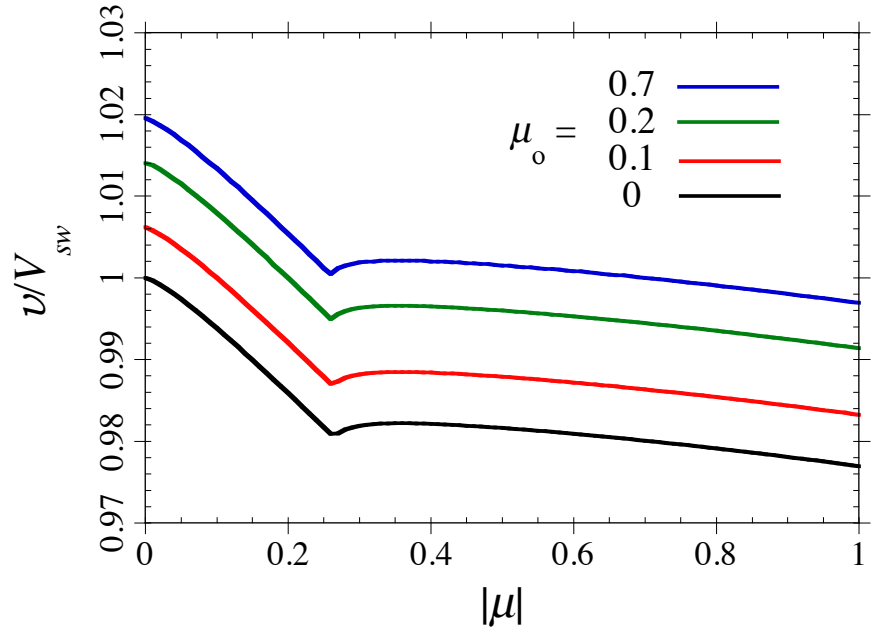

Figure 2

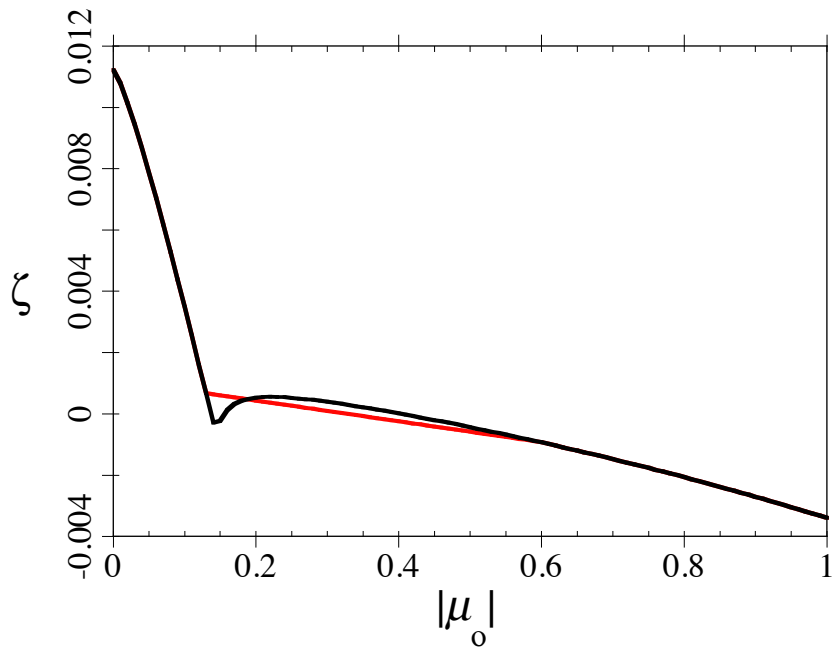

Figure 3

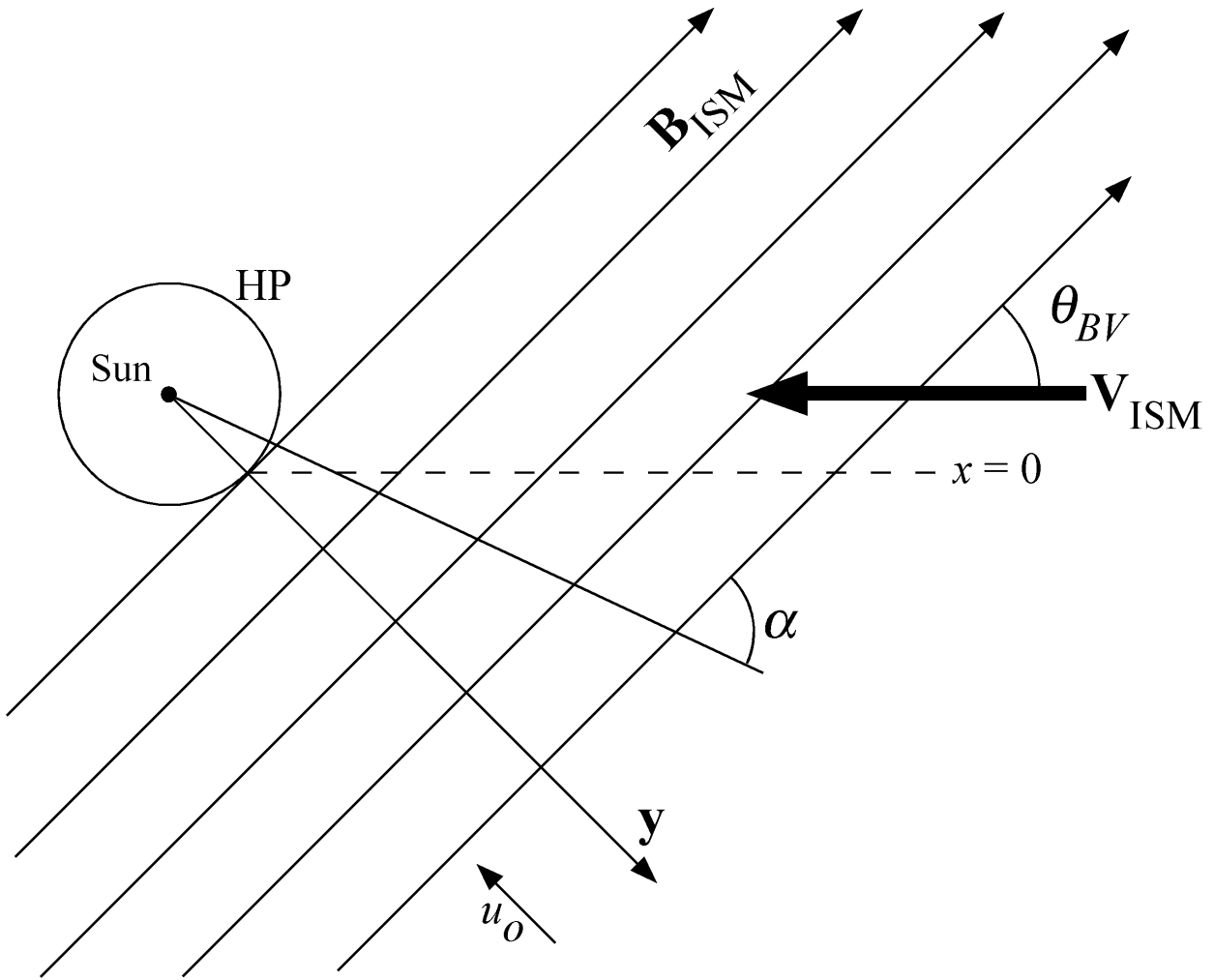

Figure 4

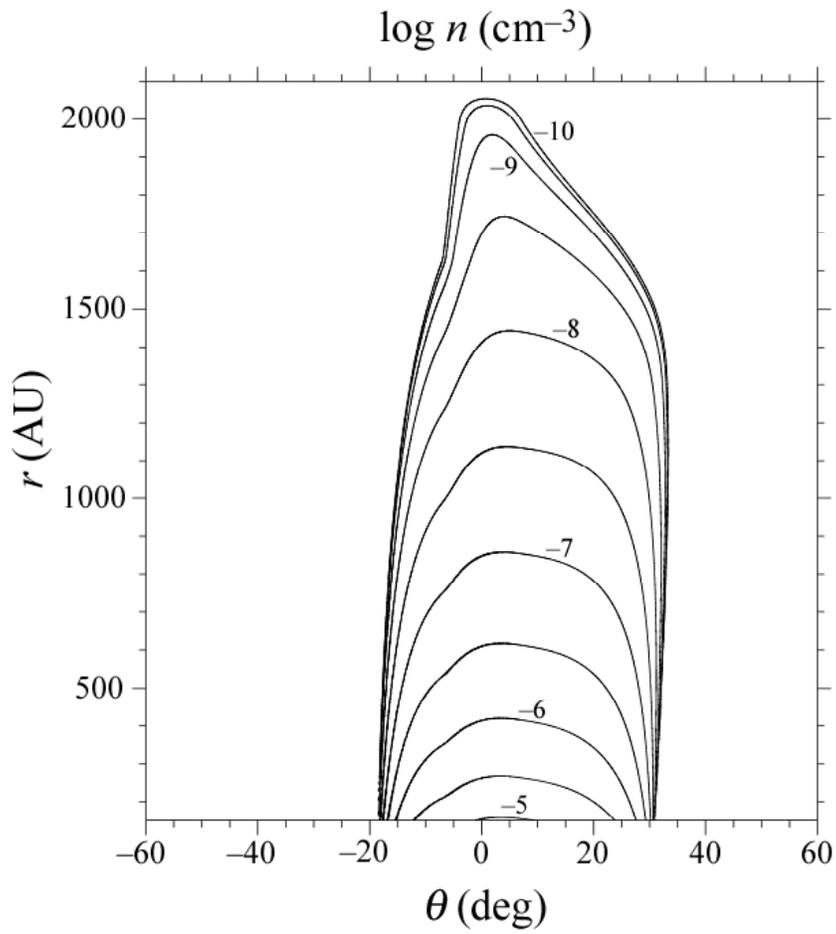

Figure 5

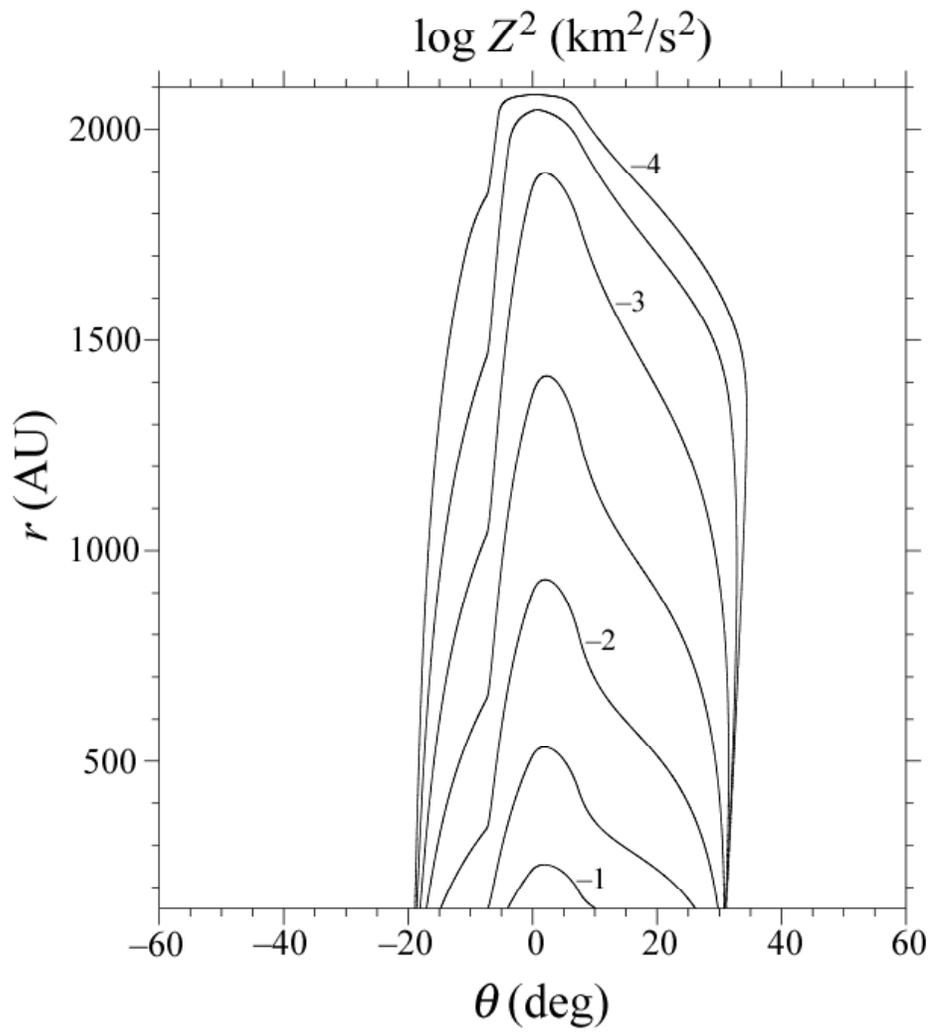

Figure 6

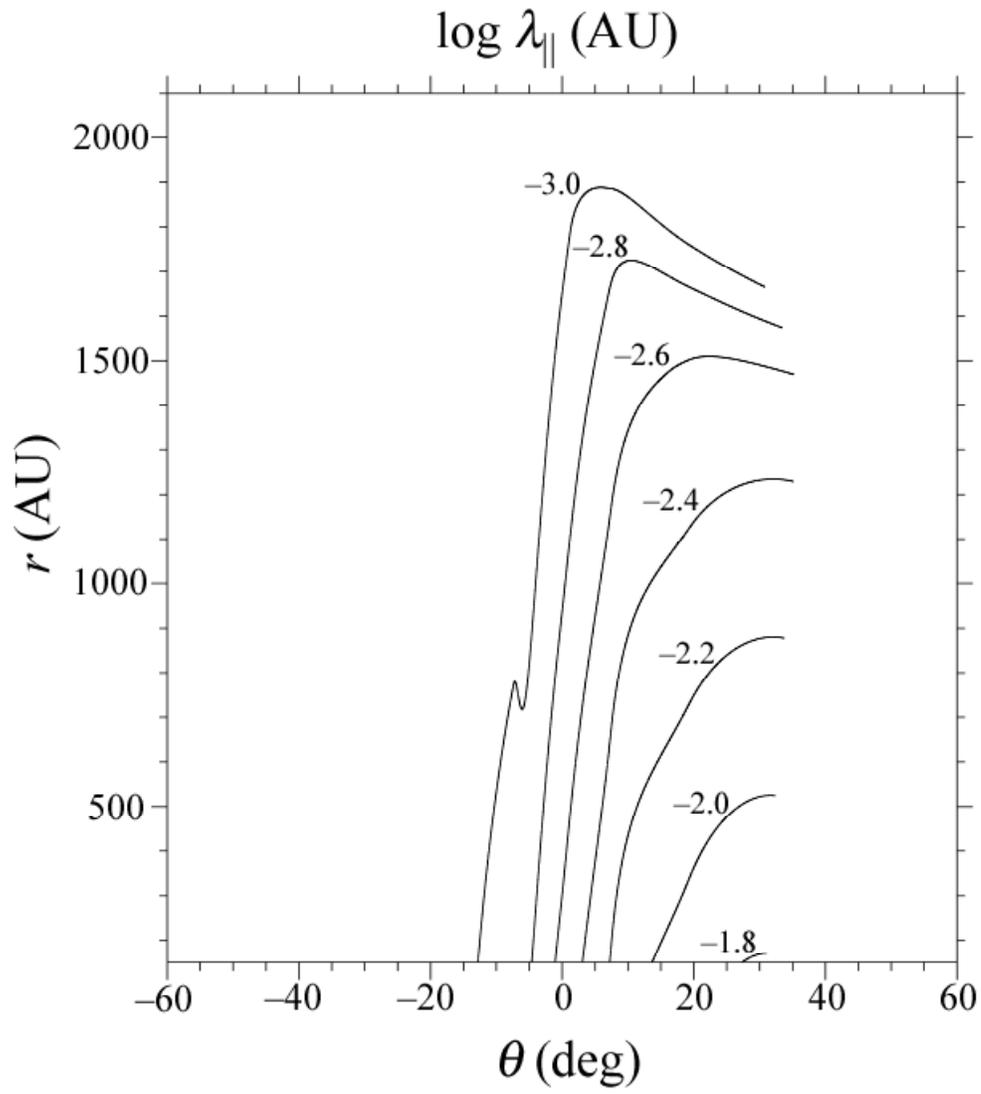

Figure 7

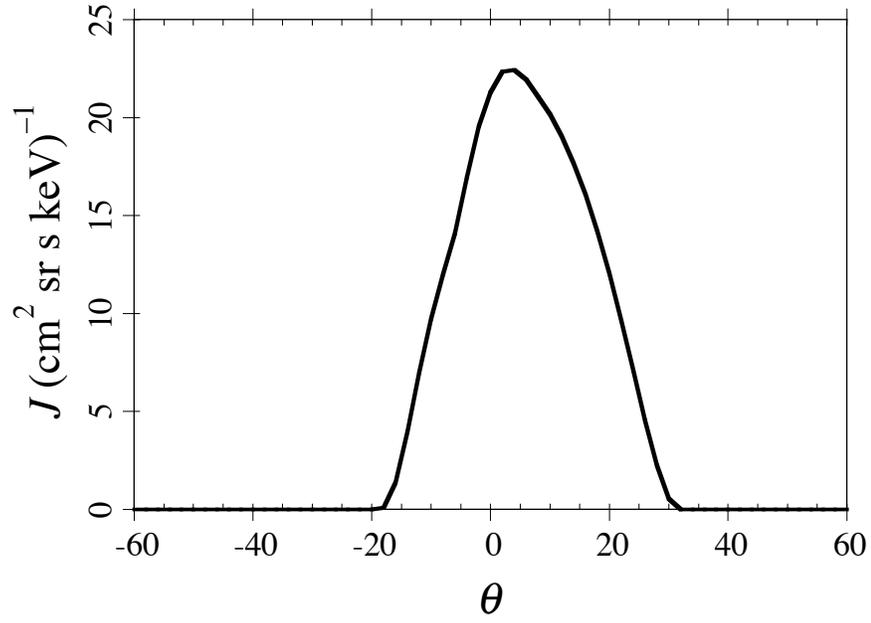

Figure 8